# Lateral Mn$_5$Ge$_3$ spin-valve in contact with a high-mobility Ge two-dimensional hole gas

*David Weißhaupt, Christoph Sürgers\*, Dominik Bloos, Hannes Simon Funk, Michael Oehme, Gerda Fischer, Markus Andreas Schubert, Christian Wenger, Joris van Slageren, Inga Anita Fischer, Jörg Schulze*

D. Weißhaupt, H. S. Funk, Dr. M. Oehme, Prof. J. Schulze
Institute of Semiconductor Engineering, University of Stuttgart, 70569 Stuttgart, Germany

Dr. C. Sürgers, Dr. G. Fischer,
Physikalisches Institut, Karlsruhe Institute of Technology, 76131 Karlsruhe, Germany
E-mail: christoph.suergers@kit.edu

Dr. D. Bloos, Prof. van Slageren
Institute of Physical Chemistry, Pfaffenwaldring 55, 70569 Stuttgart, Germany

Prof. I. A. Fischer
Experimental Physics and Functional Materials, Brandenburg University of Technology, 03046 Cottbus, Germany

Dr. M. A. Schubert, Prof. C. Wenger
Materials Research, IHP – Leibniz Institut für innovative Mikroelektronik, 15236 Frankfurt (Oder), Germany

**Abstract:**

Ge two-dimensional hole gases in strained modulation-doped quantum-wells represent a promising material platform for future spintronic applications due to their excellent spin transport properties and the theoretical possibility of efficient spin manipulation. Due to the continuous development of epitaxial growth recipes extreme high hole mobilities and low effective masses can be achieved, promising an efficient spin transport. Furthermore, the Ge two-dimensional hole gas (2DHG) can be integrated in the well-established industrial complementary metal-oxide-semiconductor (CMOS) devices technology. However, efficient electrical spin injection into a Ge 2DHG - a prerequisite for the realization of spintronic devices - has not yet been demonstrated. In this work, we report the fabrication and low-temperature magnetoresistance measurements of a laterally structured $Mn_5Ge_3$/Ge 2DHG/ $Mn_5Ge_3$ device. The ferromagnetic $Mn_5Ge_3$ contacts are grown directly into the Ge quantum well by means of an interdiffusion process with a spacing of approximately 130 nm. We observe a magnetoresistance signal for temperatures below 13 K possibly arising from successful spin injection. The results represent a step forward toward the realization of CMOS compatible spintronic devices based on a 2DHG.

# 1. Introduction

New concepts summarized under the term "Beyond complementary metal-oxide-semiconductor (CMOS)" are required in order to further increase the performance of semiconductor integrated circuits.[1,2] One prominent example is the field of semiconductor spintronics where the electron spin is used in devices in addition to its charge.[3–5] For a viable spintronic device several requirements must be fulfilled. First, spins must be successfully injected into a high-conductivity channel. Second, the spin orientation must be maintained over a long distance, requiring in turn high charge carrier mobilities and long spin relaxation times. Third, there must be a means to efficiently manipulate the spin orientation. The spin-field-effect transistor (spin FET) proposed in 1990 by Datta and Das, is a prototypical semiconductor spintronics device [6] in which spin polarized charge carriers are injected from ferromagnetic source electrodes into a semiconductor channel. Depending on the relative orientation of the spin of the charge carriers and the magnetization of the drain electrode, the transistor is in its on- or off-state. The spin of the charge carrier is switched between the parallel and anti-parallel configuration relative to the magnetization of the ferromagnetic drain contact through the gate-controlled Rashba spin-orbit interaction resulting in an oscillating output characteristic that is unique for the spin FET and is largely determined by the giant magnetoresistance (GMR) effect.[7,8]

For efficient spin manipulation, materials lacking inversion symmetry and strong spin-orbit interaction generating a sizeable Rashba effect are required.[9–12] For potential industrial applications, the spin FET is discussed as a low-power transistor, since the energy for switching the spin orientation is orders of magnitude smaller than that of the Coulomb charging energy in a classic metal-oxide-semiconductor FET (MOSFET), thereby strongly reducing the heat dissipation in conventional semiconductor devices.[13–15] From a research point of view, the spin FET is of great interest because it does not only require electrical spin injection and spin detection as well as spin transport, but also spin manipulation by means of an electrical field. The unification of all components within one device makes it an ideal demonstrator and thus at the same time the foundation for future spintronic devices. However, the implementation of this device concept has proven to be extremely difficult. In their seminal paper, Datta and Das proposed that spin transport takes place in a buried high-mobility two-dimensional *electron* gas (2DEG) [6]. Even though Lee et al. were able to demonstrate a working Si-based spin FET [16], there is only one example [17] where a non-local Rashba oscillation within an InAs 2DEG with ferromagnetic $Ni_{81}Fe_{19}$ contacts has been demonstrated. The main challenge is the fabrication of ferromagnetic contacts to the buried channel without depleting the 2DEG. While various groups have reported successful electrical spin injection into buried 2DEGs in group III-V compound semiconductors [18–23], spin injection into a Si 2DEG was successfully demonstrated in only one study [24]. Unfortunately, the spin-orbit interaction in Si is comparably weak and materials with larger spin-orbit interactions are highly desired. [25–27]

A promising and CMOS compatible material to meet all the requirements are Ge-based heterostructures forming a two-dimensional *hole* gas (2DHG). [28,29] In contrast to the 2DEG, the 2DHG provides an increased Rashba energy due to the larger spin-orbit interaction of holes compared to that of electrons giving rise to a more efficient spin manipulation.[30] On the one hand, large Rashba energies comparable to the group III-V compound semiconductors have been obtained by various methods such as weak anti-localization (WAL), magnetoresistance (MR), and cyclotron resonance measurements. [31–37] On the other hand, electronic transport is degraded by the fundamentally shorter spin relaxation time of holes compared to electrons. However, even if the exact relationship of spin-flip length on the momentum scattering is unclear, it is reasonable to assume that the spin orientation of holes is strongly linked to their momentum, so that the spin information is typically randomized with each scattering event. Consequently, the spin relaxation time of holes is of the same time scale as the transport scattering time.[38,39] While the latter is commonly affected by large-angle scattering, the quantum scattering time considers each scattering event regardless of the scattering angle and can be used as a worst-case estimate for the spin relaxation time. Due to the ongoing development of growth recipes for the Ge 2DHGs, extremely high hole-mobilities and low effective masses have already been achieved resulting in enhanced transport and quantum scattering times for holes.[40–43] Therefore, the Ge 2DHG promises a high spin relaxation time and thus very good spin transport properties. In this respect it is remarkable that despite the good suitability of the Ge 2DHG for CMOS-compatible spintronic applications, electric spin injection into a Ge DHG has not been yet reported. So far, all spin injection experiments based on p-type material have been carried out on doped Ge bulk samples with three-terminal or four-terminal (4T) structures [44–51], which fundamentally differ in transport mechanism compared to a buried Ge 2DHG channel.

Here, we report for the first time MR measurements performed on a lateral spin-valve device with a high-mobility Ge 2DHG as spin transport channel. The Ge/Si$_{1-x}$Ge$_x$ heterostructure was epitaxially grown on a (111) oriented Si substrate by molecular beam epitaxy (MBE). We used thin Mn$_5$Ge$_3$ layers as ferromagnetic contacts, which were grown directly into the buried Ge quantum-well (QW) by means of interdiffusion. To this end, the entire Si$_{1-x}$Ge$_x$ capping layer above the Ge QW was removed prior to contact formation.[52] Temperature dependent MR measurements were first carried out on Hall bar structures to extract the quantum scattering time and effective mass as characteristic parameters for the spin transport properties. The magnetic properties were then analyzed using a superconducting quantum interference device (SQUID) magnetometer. Finally, the MR measurements on the structured devices exhibit signals hinting at spin injection into the Ge 2DHG. The GMR signal depends on the selected operating-point and could be observed up to a temperature of $T = 13$ K. The results reported in this work are important for the realization of CMOS compatible spintronic devices.

## 2. Device Fabrication and Characterization

### 2.1. Electronic transport properties of the Ge 2DHG

The 2DHG was epitaxially grown on a Si (111) substrate following a standard growth protocol for the (100) substrate orientation reported earlier. [64,65] The comprehensive crystal analysis by high resolution X-ray diffraction (HR-XRD), transmission electron microscopy (TEM), atomic force microscopy (AFM) of the grown Ge 2DHG sample, as well as the fabrication of the Hall bar and GMR device are provided in the Supplemental Material.

Fig. 1 (a) shows the longitudinal MR $R_{xx}$ and the transverse Hall resistance $R_{xy}$ in an applied magnetic field of up to $\mu_0 H = 15$ T at a temperature of $T = 2$ K, from which a Hall mobility and Hall sheet carrier density of the Ge 2DHG of $\mu = (3.02 \pm 0.01) \times 10^4$ cm$^2$V$^{-1}$s$^{-1}$ and $p_{s,\text{Hall}} = (4.62 \pm 0.16) \times 10^{11}$ cm$^{-2}$, respectively, were obtained. The corresponding temperature dependent Hall measurement is shown in the Supplemental Material. Fig. 1 (b) depicts the temperature dependence of the MR for a magnetic field range from $\mu_0 H = -6$ T to $\mu_0 H = 6$ T. Clear Shubnikov-de Haas (SdH) oscillations and integer quantum Hall plateaus were observed, highlighting the quality of the epitaxially grown layers and therefore the good transport properties of the Ge 2DHG. The SdH oscillations start at approximately $\mu_0 H = 0.65$ T, with spin splitting occurring at $\mu_0 H = 3.13$ T. This is an indication for strong spin-orbit interaction and thus the possibility for efficient spin manipulation. Furthermore, the clear separation of the Landau levels for high magnetic fields gives further proof of the excellent transport properties of our samples. The symmetry of the MR curve excludes the presence of any inhomogeneities caused by, e.g., fluctuations in modulation doping.[53]

A sheet carrier density $p_{s,\text{SdH}} = (4.00 \pm 0.30) \times 10^{11}$ cm$^{-2}$ was obtained from the period of the SdH oscillation and thus slightly deviates from the Hall measurement results. This difference indicates the presence of parasitic channels, which could either originate from the modulation-doped layer or from the Si substrate, which is relatively highly doped. However, since the relative amount of charge carriers outside the Ge 2DHG is quite small, the following GMR measurements are not affected.

From the temperature dependent damping of the amplitude of the SdH oscillations an effective mass of $m^* = (0.084 \pm 0.001) \times m_0$ ($m_0$: electron mass) and a quantum-scattering time of $\tau_q = (0.45 \pm 0.01)$ ps was extracted. At this point, we can estimate the spin-flip length $l_{sf}$ using our experimentally determined transport data. In a worst-case scenario, we assume that the spin-information in the Ge 2DHG is randomized within each scattering event, i.e., the expected spin relaxation time $\tau_{sf}$ corresponds to the quantum scattering time $\tau_q$. For the associated lower limit of the spin-flip length we then get $l_{sf} = \sqrt{D\tau_q} \approx (133.84 \pm 0.01)$ nm, where $D = (395.2 \pm 4.7)$ cm$^2$/s is the diffusion constant of the holes in the Ge 2DHG.

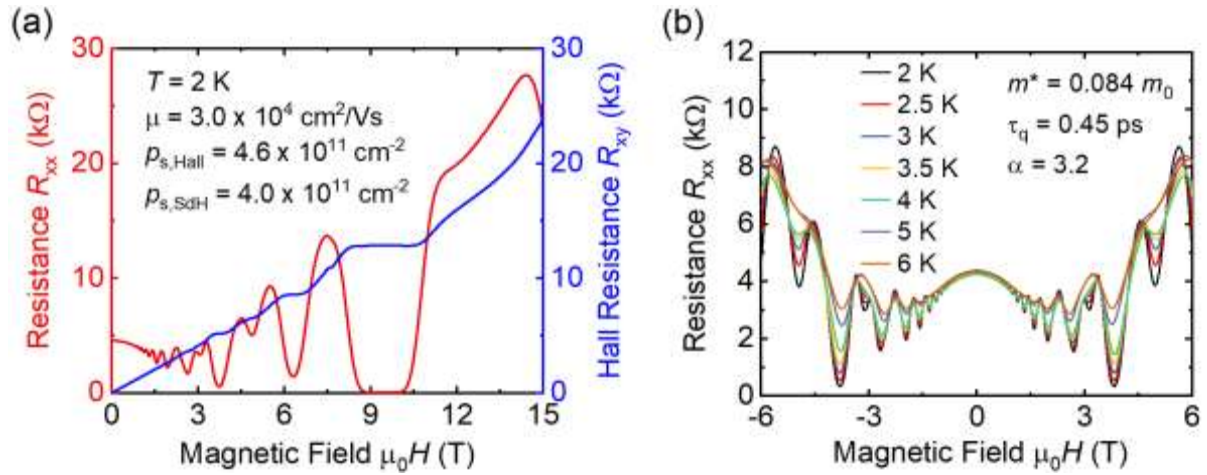

**Figure 1.** Transport properties of the Ge 2DHG: (a) Longitudinal resistance $R_{xx}$ and transverse Hall resistance $R_{xy}$ vs magnetic field $H$ at $T = 2$ K. (b) Temperature dependence of $R_{xx}$.

## 2.2. Ferromagnetic Mn$_5$Ge$_3$ contacts

The ferromagnetic Mn$_5$Ge$_3$ contacts were formed directly in the Ge QW using an interdiffusion process. Prior to the Mn$_5$Ge$_3$ contact formation, the entire Si$_{1-x}$Ge$_x$ capping layer on top of the Ge QW was removed using an Ar$^+$ ion milling process. For a detailed description of the fabrication process and a discussion of the magnetic properties of an unstructured reference sample we refer to existing literature.[52,54] The ferromagnetic Mn$_5$Ge$_3$ contacts in this work were encapsulated with an additionally evaporated Mn protection layer. For the detection of electrical spin injection, we used two different lateral sizes of the Mn$_5$Ge$_3$ contact: $A = 5$ µm × 20 µm and $B = 10$ µm × 20 µm. The Mn$_5$Ge$_3$ contacts were analysed by high-resolution transmission electron microscopy (TEM) and energy-dispersive X-ray spectroscopy (EDX) using a Tecnai Osiris electron microscope from FEI operated at an acceleration voltage of 200 kV. EDX was performed in scanning mode using the software "Esprit" from Bruker.

Fig. 2 (a) depicts the schematic cross-section of a single $Mn_5Ge_3$ contact and Fig. 2 (c) the corresponding scanning TEM (STEM) and EDX images confirming that the $Mn_5Ge_3$ contact formed upon annealing is located directly within the Ge QW. The fabrication process leads to the formation of an additional $Mn_5(Si_{1-x}Ge_x)_3$ layer within the $Si_{1-x}Ge_x$ capping layer along the etched sidewall which creeps up to approximately $l = 20$ nm underneath the $SiO_2$ hard mask. However, these $Mn_5(Si_{1-x}Ge_x)_3$ layers do not create a short circuit below the $SiO_2$ layer between the center contacts and they also do not affect the contact separation distance since only the Ge 2DHG is conductive at low temperatures.

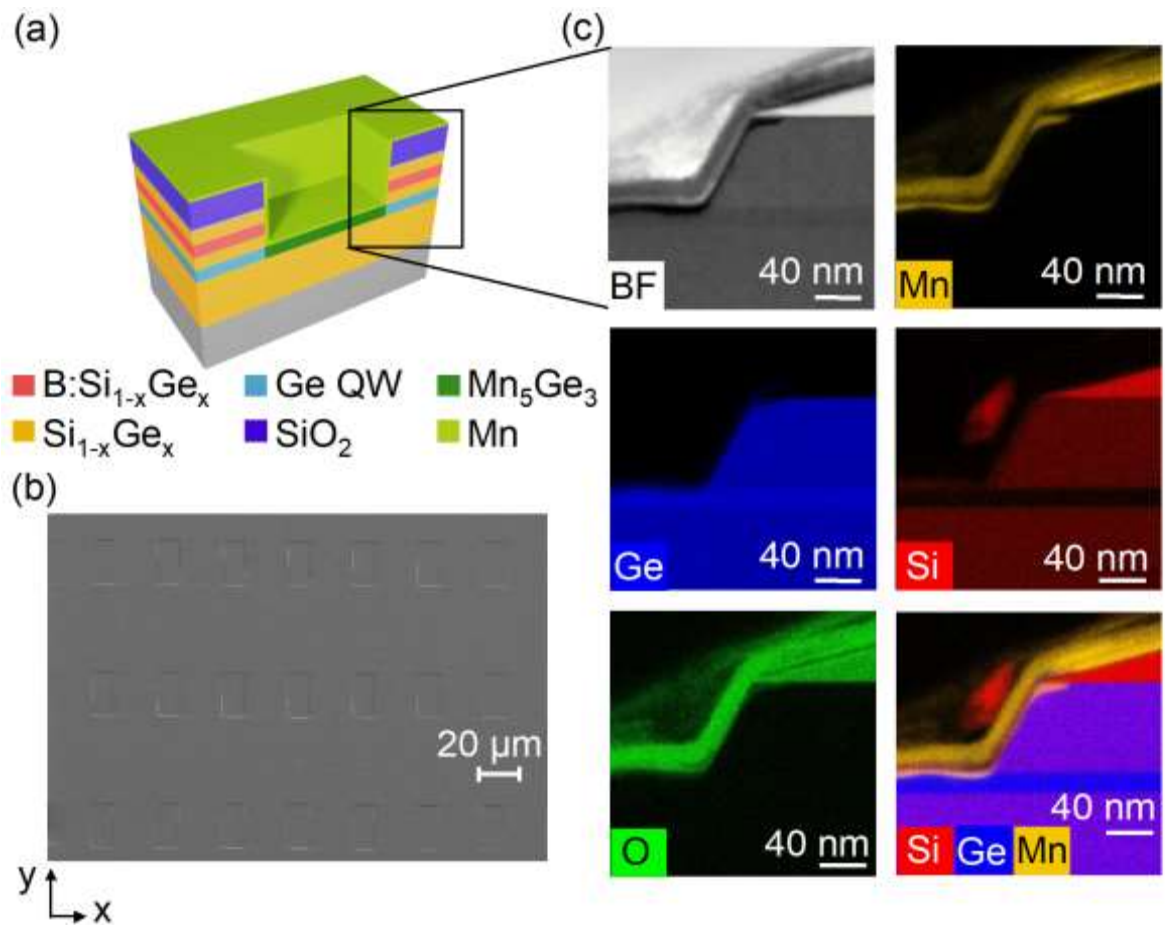

**Figure 2.** Structural properties of the $Mn_5Ge_3$-contacts: (a) schematic cross-section of a single $Mn_5Ge_3$ contact. Further details are given in the Supplemental Material. (b) Top-view SEM-image section of the $Mn_5Ge_3$ contact array with a single $Mn_5Ge_3$ contact size of $A_1 = 10$ µm $\times$ 20 µm. (c) Cross-sectional TEM and EDX images of a single $Mn_5Ge_3$ contact.

Fig. 2 (b) depicts the top-view scanning electron microscope (SEM) image of a $Mn_5Ge_3$ contact array with a single $Mn_5Ge_3$ contact size of $A_1 = 10$ µm $\times$ 20 µm. For the magnetization measurements, the Mn$_5$Ge$_3$ contacts with the two different geometries ($A_1$ = 10 µm × 20 µm, $A_2$ = 5 µm × 20 µm,) were arranged as an array extended over an area of 3 mm × 3 mm and their data were compared with data obtained from an unstructured reference sample.

The magnetization measurements were performed with a superconducting quantum interference device (SQUID) magnetometer (MPMS, Quantum Design) with the *y* axis of the contacts, see Fig. 2 (b) orientated along the in-plane external magnetic field. Fig. 3 (a) shows the temperature dependent in-plane magnetization for the different geometries in a magnetic field µ$_0$H = 50 mT. For comparison, the magnetization curves are normalized with respect to their magnetic moment at a temperature of $T$ = 5 K. For all geometries, the temperature-dependent magnetization confirms the formation of ferromagnetic Mn$_5$Ge$_3$ layers. Independent of the geometry, we obtain a Curie temperature of about 300 K. Among the possible compounds of Mn and Ge that can form, only Mn$_5$Ge$_3$ is ferromagnetic with a Curie temperature around $T$ = 300 K and exhibits no other magnetic phase transition.[55,56] However, we observe a strong change of the slope of the magnetization at a temperature of about $T$ = 40 K, which arises from the Mn$_5$(Si$_{1-x}$Ge$_x$)$_3$ layer that was additionally generated within the Si$_{1-x}$Ge$_x$ capping layer along the etched sidewall.[57] The respective magnetization curves measured at a temperature of $T$ = 5 K are shown in Fig. 3 (b). The inset depicts the same data up to maximum external field to µ$_0$H = 3 T. The curves were corrected for a diamagnetic background signal arising from the semiconductor substrate and the sample holder. Again, for comparison all magnetization curves are normalized with respect to their saturation magnetic moments. We observe a double hysteresis at $T$ = 5 K, which again can be explained by the overlay of the Mn$_5$Ge$_3$ layer with the Mn$_5$(Si$_{1-x}$Ge$_x$)$_3$ layer. The obtained coercive fields are µ$_0$H$_C$ = (116 ± 10) mT, µ$_0$H$_C$ = (46 ± 10) mT, and µ$_0$H$_C$ = (73 ± 10) mT for the geometries of $A_2$ = 5 µm × 20 µm, $A_1$ = 10 µm × 20 µm, and the unstructured reference sample, respectively. The direct comparison of the two geometries $A_1$ and $A_2$ indicates magnetic hardening, i.e., an increase in coercivity for the $A_2$ sample which is in line with the general increase of the intrinsic coercivity with decreasing particle size for multidomain particles. From a magnetic point of view, the ferromagnetic Mn$_5$Ge$_3$ contacts meet the magnetic requirements for the detection of electrical spin injection and thus can be used for further experiments.

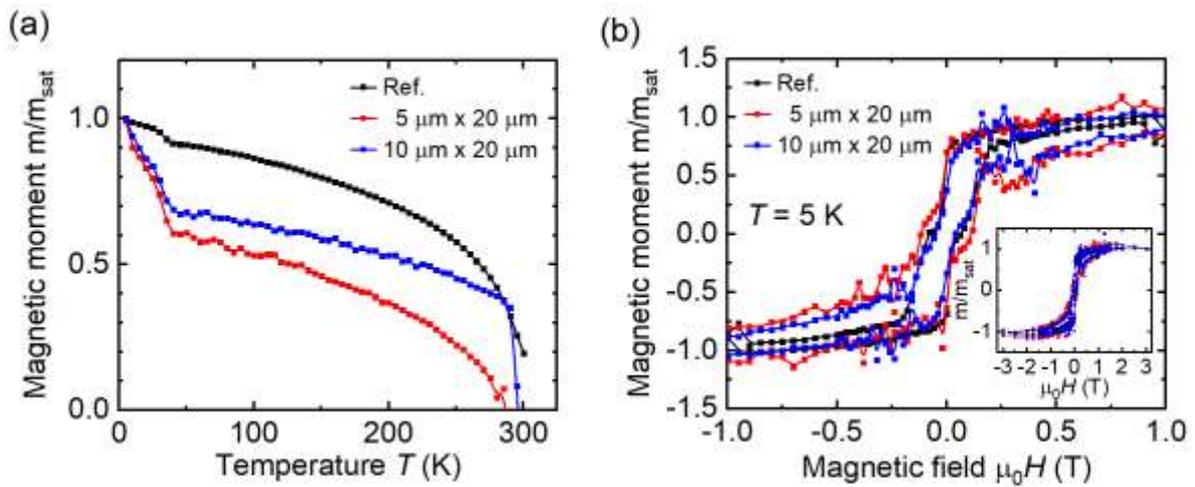

**Figure 3.** Magnetic properties of the $Mn_5Ge_3$ contacts with different geometries: (a) temperature dependent in-plane magnetization measured in an external magnetic field of $B = 50$ mT. (b) Corresponding in-plane magnetization curves at $T = 5$ K. For both measurements the external in-plane magnetic field was orientated along the $y$ axis, see Figure 3 (b). Black curves represent data from the unstructured reference sample ``Ref.´´.

## 2.3. Measurements on the GMR device

Fig. 4 gives a schematic overview of the GMR device with spin transport in the Ge 2DHG taking place between two ferromagnetic $Mn_5Ge_3$ contacts.

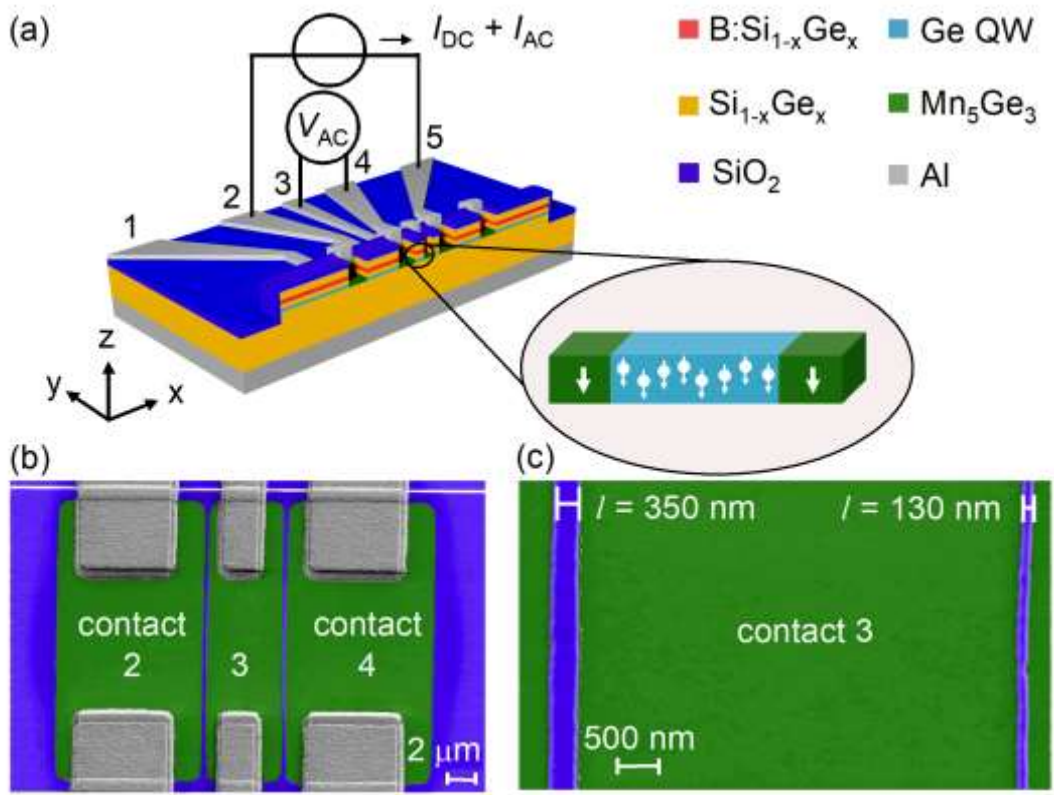

**Figure 4.** (a) Schematic illustration of the device structure used for the GMR measurements in this study. The spins are electrically injected and detected using ferromagnetic $Mn_5Ge_3$ contacts. The spin transport takes place in the Ge QW which accommodates a high mobility 2DHG. (b) and (c) show the top-view SEM images of the surface with different magnifications.

The MR measurements on the GMR device were performed in a Physical Property Measurement System (PPMS, Quantum Design) for various temperatures and magnetic fields oriented in the plane of the sample and parallel to the long side of the $Mn_5Ge_3$ electrodes, i.e., along the y axis (Fig. 4). A DC voltage $V_{DC}$ was modulated by an AC voltage $V_{AC}$ of frequency $f_{ref} = 321.7$ Hz and amplitude of a few percent of $V_{DC}$ applied to the device under test, generating currents $I_{DC}$ and $I_{AC}$, in series with a resistor $R_{ref} = 1$ kΩ and an *I-V* converter (FEMTO DLCPA200) with a gain of $10^3$. The current $I_{DC}$ through the device was determined from the voltage drop $V_{ref}$ across $R_{ref}$ measured with a Keithley 2182 voltmeter. The differential resistance $dR = dV/dI$ was determined by two lock-in amplifiers (Standford Research Systems SRS 830) synchronously coupled to $f_{ref}$ and measuring the voltage $dV$ across the device and a signal proportional to $dI$ at the output of the *I-V* converter. In addition, the DC resistance was determined by the voltage drop across the device measured with a Keithley 2182 voltmeter divided by $I_{DC}$.

## 3. Results and Discussion

The device comprises four ferromagnetic $Mn_5Ge_3$ electrodes in contact with the semiconducting Ge 2DHG transport channel, with spin transport taking place between the two center contacts separated by a distance lower than the spin-diffusion-length. A detailed description of the fabrication process of the device is available in the Supplemental Material. Fig. 4 shows the schematic cross section of the final device together with top-view SEM images of the inner contacts (contacts 2, 3, and 4) with different magnification. The $Mn_5Ge_3$ contacts formed by the interdiffusion process act as barriers in the Ge channel for the 2DHG. Therefore, the standard non-local measurement approach where the current is applied between contacts 2 and 3 and the voltage is measured between contacts 4 and 5 (see Fig. 4) to verify spin injection, cannot be performed with the present device. Hence, we follow the same approach as adopted in Ref. [24,58], where superimposed direct (DC) and alternating currents (AC) are applied between the outer contacts (contact 2 and 5) and the AC voltage between the inner contacts (contact 3 and 4) is measured. The DC current serves to bring the device to a suitable operation point (see below) and the AC current serves for the actual measurement.

Furthermore, the final design of the device differs from the well-established 4T-Hanle structure in that five contacts are used instead of four allowing spin-transport measurements over two different distances within one device. The two distances between the center electrode and the outer right or left electrode are approximately $l = 130$ nm and $l = 350$ nm, respectively, see Fig. 4. The size of the $Mn_5Ge_3$ center contact is $A_2 = 5$ µm $\times$ 20 µm, all other contacts are twice as wide with a size of $A_1 = 10$ µm $\times$ 20 µm.

Fig. 5 (a) shows MR curves $dR = dV/dI$ of the $Mn_5Ge_3$/Ge 2DHG/$Mn_5Ge_3$ device with a channel length of $l = 130$ nm at a temperature of $T = 2$ K for different DC currents $I_{DC}$. While no signal was measured for small DC currents, a clear GMR signal develops as the DC current $I_{DC}$ increases. For $I_{DC} = 24$ µA and $V_{DC} = 6$ V, corresponding to a resistance $R = 250$ kΩ, a $dR = 8.9$ kΩ and a MR of $dR/R = 3.6\%$ were obtained. A dependence of the electric spin injection on the applied DC current has already been reported for configurations involving $Mn_5Ge_3$/Ge [58] and $Mn_5(Si_{1-x}Ge_x)_3$/Si [24] contacts and has been attributed to the fact that the $Mn_5Ge_3$ contact on the Ge surface forms a Schottky barrier with a barrier height of $\phi_B = 0.25$ eV [59] which blocks electric spin injection. When a sufficiently high bias is applied to the Schottky contact, spin polarized holes can tunnel through the Schottky barrier, which becomes narrower at higher bias. This leads to the formation of spin polarized currents in the Ge 2DHG. The spin polarized electrons then migrate in the Ge 2DHG to the second ferromagnetic contact. During this process the spin polarization decays exponentially due to scattering. The relative orientation between the spins in the Ge 2DHG and the magnetization of the second $Mn_5Ge_3$ contact then defines the total resistance and thus the signal structure.

Sweeping the external magnetic field leads to switching of the magnetization between the parallel and anti-parallel configurations of the two ferromagnetic $Mn_5Ge_3$ contacts. For large external magnetic fields, the magnetization of the two $Mn_5Ge_3$ contacts follows the direction of the external field, i.e., their mutual orientation is parallel. When the direction of the magnetic field changes in the reversed direction, the different coercive fields of the two electrodes cause the magnetization in the contacts to switch at different absolute field strengths. In our sample, the low spatial separation between the middle ferromagnetic contacts can be expected to result in a sizeable antiferromagnetic coupling between neighbouring contacts due to dipole-dipole interactions. Our MR signal, therefore, shows a different dependence on the external magnetic field than for 4T-Hanle structures with larger contact spacing, which usually exhibits GMR behaviour with separated MR hysteresis. When sweeping the amplitude of the in-plane external magnetic field from large positive values to large negative values, we obtain a change in our MR signal already at small positive values of the external field instead of at small negative values, indicating anti-parallel orientation of the magnetization in the neighbouring contacts prior to the sign change of the magnetic field.

The GMR signal is shifted towards negative external fields by about $\Delta\mu_0 H = (-20 \pm 10)$ mT. Such shifting is typically caused by a coupling of the ferromagnet with an antiferromagnetic thin film, known as the exchange-bias effect.[60] In our samples, the generation of local changes in alloy composition during the annealing step for the formation of the Mn-based ferromagnetic contacts could lead to this effect.[57] Indeed, the magnetic hysteresis curves for the geometries of $A_2 = 5$ µm $\times$ 20 µm, $A_1 = 10$ µm $\times$ 20 µm, and the unstructured reference sample, presented in Fig. 3 (b), indicate the presence of a small shift in magnetic field by approximately $\Delta\mu_0 H = (-5 \pm 10)$ mT, $\Delta\mu_0 H = (-13 \pm 10)$ mT, and $\Delta\mu_0 H = (-19 \pm 10)$ mT, respectively. However, it is more likely that the strongly enhanced demagnetization fields in microstructured samples may also lead to an antiferromagnetic coupling and a strongly field-shifted GMR behaviour.[61] A similar behaviour has been observed for lateral 2-terminal structures on the $Si_{0.1}Ge_{0.9}$ platform when sweeping a minorloop.[62] Hence, independent of the origin of the shifted GMR characteristics the DC current dependent GMR signals provide strong evidence for successful spin injection, transport, and detection in the Ge 2DHG.

Fig. 5 (b) shows the GMR signal for different temperatures. For this purpose, the operating point $I_{DC}$ was chosen so that the relative signal for the respective temperature is at its maximum. The signal is significantly attenuated with increasing temperature, which is in good agreement with the results of the magnetotransport measurements, i.e., the SdH oscillations are strongly damped in the range from $T = 2$ K to $T = 6$ K, indicating an increase in scattering with increasing temperature.

Overall, the GMR signal could only be observed for a short distance of $l = 130$ nm between the electrodes. The sample with a larger electrode distance of $l = 350$ nm shows no evidence of spin injection (not shown), which means that the associated spin-flip length is expected to be somewhere between these two distances. Thus, the lower limit spin-flip length of about $l_{sf} = (133.84 \pm 0.01)$ nm calculated on the basis of the quantum-scattering time is in good agreement with the experimental results presented here.

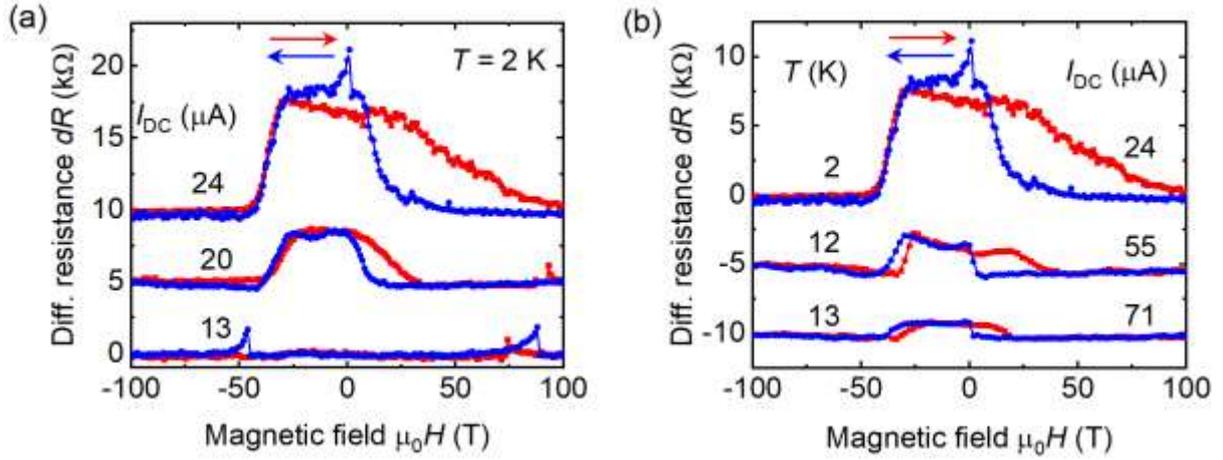

**Figure 5.** MR measurements of the $Mn_5Ge_3$/ Ge 2DHG/$Mn_5Ge_3$ device with a channel length of $l$ = 130 nm. (a) GMR signal for different DC currents $I_{DC}$ at $T$ = 2 K. (b) GMR signal at different temperatures with an adjusted operating point $I_{DC}$. For the sake of clarity, a constant MR has been removed from the raw data and the measurements are shifted with respect to each other.

The MR measurements presented in Fig. 5 were performed at a fixed sweep rate of the external magnetic field of $R$ = 0.2 mT/s. However, our experiments have shown that the sweep rate has a decisive influence on the signal structure. Fig. 6 compares the MR-curves for different sweep rates of the external magnetic field. The measurements shown were carried out at a temperature $T$ = 2 K with an applied DC current of $I_{DC}$ = 25 µA. The GMR signal only appears for sweep rates below $R$ < 1.0 mT/s, whereas no signal structure can be observed for high sweep rates of $R$ = 5.0 mT/s. MR measurements with sweep rates in the range of 1.0 mT/s $\leq R \leq$ 5.0 mT/s show signal structures resembling spin-valve signals. This dependence on the sweep rate of the external magnetic field results from the switching dynamics behavior of the ferromagnetic $Mn_5Ge_3$ contacts. The alignment of the magnetization or the individual domains in the external magnetic field does not take place instantaneously, but is also retarded with a certain time constant. Sweep rates that are too fast lead to a modification of the GMR signal, since the individual domains cannot follow the rapidly changing external magnetic field. This influence of the magnetic field sweep rate on magnetic transitions has already been observed for synthetic ferrimagnets with large lateral size and could also explain our observation. Thus, modifying the sweep rate leads to change of the magnetization reversal process from domain wall propagation to nucleation, the latter dominating at a high sweeping rate.[63] The spin-valve-like signal for 2.0 mT/s therefore corresponds to a GMR signal which shifts on the magnetic field axis with decreasing sweep rate until both measurements match. For sweep rates higher than $R$ = 5.0 mT/s, there is no longer any defined magnetic switching behavior, which is why no change in resistance can be detected electrically.

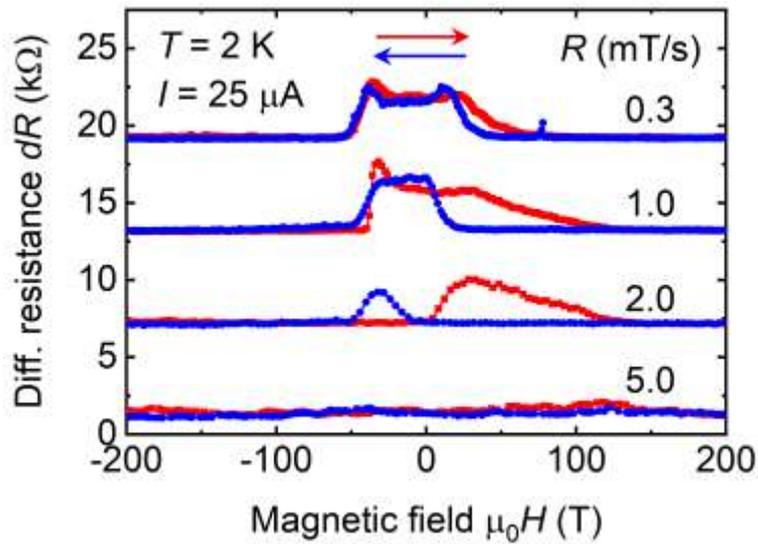

**Figure 6.** MR measurements for different sweep rates $R$ of the external magnetic field at $T = 2$ K and for an applied DC current is $I_{DC} = 25$ µA. A constant MR has been removed from the raw data and the various measurements are shifted for clarity.

## 4. Conclusion

Low-temperature magnetoresistance measurements performed on a lateral ferromagnet/Ge 2DHG/ferromagnet device structure show a clear GMR signal for a contact distance of 130 nm between the electrodes. To this end, we use ferromagnetic $Mn_5Ge_3$ contacts which were directly grown as a thin film into the Ge QW. Depending on the DC current $I_{DC}$, we observe GMR signals up to a temperature of $T = 13$ K. The attenuation of the GMR signal with increasing temperature is in agreement with the quantum-mechanical transport properties of the Ge 2DHG. These are important results for the realization of future CMOS compatible spintronic devices, in particular for devices based on the spin FET proposed by Datta and Das.

**Supplemental Material**

See Supplemental Material at [URL will be inserted by publisher] for structural analysis, temperature dependent Hall measurements, and fabrication of the mesa.

**Acknowledgements**

We acknowledge funding by the Deutsche Forschungsgemeinschaft (DFG) under Grants FI 1511/3-1 & SU 225/3-1 as well by the Center for Integrated Quantum Science & Technology (IQST). D. B. and J. v. S. thank the Ministry of Science, Research and Art Baden-Württemberg (MWK) as well as the University of Stuttgart for funding (Research Seed Capital).